\documentclass[prb,a4paper,twocolumn,floatfix]{revtex4-1}
\usepackage{natbib}
\usepackage{amssymb,amsthm,amsmath, amscd}
\usepackage[T1]{fontenc}
\usepackage[ansinew]{inputenc}
\usepackage{times}
\usepackage{graphicx}
\usepackage{bm}
\begin{document}

\title{Quasi-non-local gradient-level exchange-correlation approximation for metals and alloys}

\author{H. Lev\"am\"aki}
\email{henrik.levamaki@utu.fi}
\author{M.P.J. Punkkinen}
\author{K. Kokko}
\affiliation{Department of Physics and Astronomy, University of Turku, FI-20014 Turku, Finland}
\affiliation{Turku University Centre for Materials and Surfaces (MatSurf), Turku, Finland}
\author{L. Vitos}
\email{levente@kth.se}
\affiliation{Applied Materials Physics, Department of Materials Science and Engineering, Royal Institute of Technology, SE-10044 Stockholm, Sweden}
\affiliation{Department of Physics and Astronomy, Division of Materials Theory, Uppsala University,
Box 516, SE-751210, Uppsala, Sweden}
\affiliation{Research Institute for Solid State Physics and Optics, \\Wigner Research Centre for Physics, H-1525
Budapest, P.O. Box 49, Hungary}

\date{02 July 2012}

\begin{abstract}
The flexibility of common generalized gradient approximation for the exchange-correlation energy is investigated by monitoring the equilibrium volume of transition metals. It is shown that no universal gradient-level approximation yielding consistent errors for all metals exists. Based on an element-specific optimization, the concept of quasi-non-local gradient-level approximation is introduced. The strength of the scheme is demonstrated on several transition metal alloys.
\end{abstract}


\maketitle

Density functional theory (DFT)\cite{Hohenberg1964} has become a fundamental first-principles research tool within the modern materials science. The great breakthrough during the last 30-40 years should primarily be attributed to the local density formalism,\cite{Kohn1965} where the many-body interactions are efficiently incorporated within an effective one-electron local potential.

The first description of the unknown exchange-correlation potential was provided by the local density approximation (LDA), assuming uniform electron density on the scale of the exchange-correlation hole. This seemingly rough approximation turned out to be unexpectedly accurate in total energy calculations, which in fact made the early DFT success possible. Attempts to go beyond LDA within the framework of the local density formalism have led to the elaboration of the density gradient corrected functionals. Accordingly, two major gradient-level density functional (GDF) families emerged. The generalized gradient approximation (GGA)\cite{Perdew1986,Perdew1996,Zhang1998,Perdew2008} aimed to stabilize the diverging term from the second order gradient expansion,\cite{Kohn1965} and gave for the first time a proper description of many important solids, such as the ferromagnetic iron. The subsystem functional approach (SFA) \cite{Armiento2002} originated from the nearsightedness principle \cite{Kohn1996} and incorporates inhomogeneous electron density effects through well adapted model systems. The simplest SFA was put forward within the Airy gas approximation,\cite{Kohn1998} which was later further developed into various gradient-level approximations.\cite{Vitos2000,Armiento2005,Constantin2009} For both GDF families, LDA represents the lowest order approximation and thus the correct limit in systems with densities showing negligible inhomogeneities.

A GDF exchange-correlation energy depends on the electron density $n$ and its gradient $\nabla n$, viz.
\begin{equation} \label{eq:Exc_gga}
E^{\rm GDF}_{\rm xc}[n] = \int {\rm d}^3r f_{xc}(r_s,s),
\end{equation}
with $r_s=(3/(4\pi n))^{1/3}$ and $s=|\nabla n|/(2n(3\pi^2n)^{1/3})$. The exchange-correlation energy density $f_{xc}(r_s,s)$ is commonly expressed with the help of the enhancement function $F_{xc}(r_s,s)$ over the exchange energy density of the uniform electron gas, $\epsilon^{\rm LDA}_x(n)$. For slowly varying electron density ($s\rightarrow 0$), $F_{xc}(r_s,s)\rightarrow F_{xc}^{\rm LDA}(r_s) = 1+\epsilon^{\rm LDA}_c(n)/\epsilon^{\rm LDA}_x(n)$ where $\epsilon^{\rm LDA}_c(n)$ is the LDA correlation. The explicit form of $F_{xc}(r_s,s)$ contains all information about the actual approximation. A detailed comparison of different enhancement functions from the GGA and SFA families\cite{Delczeg2011} shows that the behavior of $F_{xc}(r_s,s)$ in the rapidly varying density regime (corresponding to $s\gtrsim 0.5$ for $r_s\lesssim 4$ and $s\gtrsim 0.2$ for $r_s\gtrsim 4$) determines the performance of the approximation for a particular inhomogeneous electron system.

In this Letter, first we investigate the flexibility of the GDFs by monitoring their behavior for bulk transition metals. Based on the SFA concept, next we introduce a quasi-non-local gradient-level approximation (QNA) that performs equally well for mono-atomic systems as well as for multi-component solid solutions. For sake of transparency, we limit the present assessment to  two familiar GGA exchange-correlation density functionals, namely the Perdew-Burke-Ernzerhof (PBE)\cite{Perdew1996} and the revised PBE (PBEsol)\cite{Perdew2008} functionals. Nevertheless, the main conclusion and the emerging QNA can easily be extended to any GDF type of description.

The PBE and PBEsol approximations are based on similar enhancement functions. The prior was designed to provide accurate atomic energies whereas the latter was optimized for bulk and surface systems by restoring the original gradient expansion behavior for the exchange part and adjusting the correlation term using the jellium surface exchange-correlation energies obtained at meta-GGA level.\cite{Tao2003} The parameters controlling these effects are $\mu$ and $\beta$ and their values for PBE and PBEsol are $(\mu,\beta)_{\rm PBE}=(0.21951,0.066725)$ and $(\mu,\beta)_{\rm PBEsol}=(0.123457,0.046000)$, respectively. The large ($44$ and $31\%$) difference between the two sets of parameters indicates that the atomic and bulk regimes require rather different enhancement functions. The LDA functional is recovered for $(\mu,\beta)_{\rm LDA}=(0,0)$.

We demonstrate the performance of different GDFs' by calculating the equilibrium volumes of seven transition metals (V, Fe, Cu, Nb, Pd, W, Au) and four alloys (V-W, V-Fe, CuAu and Cu$_3$Au). The electronic structure and total energy calculations were performed using the exact muffin-tin orbitals method.\cite{Andersen1994,Vitos2000a,Vitos2001} The Kohn-Sham equations were solved within the scalar relativistic approximation, the Green's function was calculated for 16 complex energy points distributed exponentially on a semicircular contour including the valence states and employing the double Taylor expansion approach.\cite{Kissavos2007} The basis set included \emph{s}, \emph{p}, \emph{d}, and \emph{f} orbitals and the core states were recalculated after each iteration. We used 285, 240, 288 and 455 inequivalent $\vec{k}$ points in the irreducible wedge of the body centered cubic (bcc), face centered cubic (fcc), L$1_0$ and L$1_2$ Brillouin zones, respectively. The random alloys (V-W, V-Fe) were treated by the coherent potential approximation.\cite{Soven1967} The equilibrium Wigner-Seitz radii ($w$) were extracted from the equation of state described by a Morse function \cite{MorseEOS} fitted to the total energies calculated for 17 different volumes around the equilibrium. All self-consistent calculations were carried out within the LDA and the gradient terms were included in the total energy within the perturbative approach.\cite{Asato1999} The average error ($\sim7\times10^{-4}$ Bohr in $w$) due to the above approximation is below the numerical accuracy of our calculations.\cite{Ropo08}

\begin{table}
\caption{\label{table1} Equilibrium Wigner-Seitz radii (in units of Bohr) at 0 K for a selected set of transition metals. Experimental values are the same as those used in Ref. \onlinecite{Staroverov2004} (Cu, Pd), Ref. \onlinecite{Jiang2003} (Fe) and Ref. \onlinecite{Mattsson2008} (Au, W). For Nb and V, the experimental values were obtained by extrapolating the room temperature lattice constants \cite{Wyckoff1963} to 0 K using the experimental thermal expansion coefficients \cite{Kaye1993} and the Debye temperature.\cite{Ho1974}}
\begin{ruledtabular}
\begin{tabular}{llll}
Element & $w_{\rm PBE}$ & $w_{\rm PBEsol}$ & $w(\rm expt)$\\ \hline
V & 2.789 & 2.752 & 2.805\\
Fe & 2.641 & 2.600 & 2.661 \\
Cu & 2.687 & 2.638 & 2.661\\
Nb & 3.080 & 3.040 & 3.066\\
Pd & 2.923 & 2.873 & 2.866\\
W & 2.970 & 2.942 & 2.940\\
Au & 3.084 & 3.028 & 3.002
\end{tabular}
\end{ruledtabular}
\end{table}

\begin{figure}
 \includegraphics[width=0.30\textwidth]{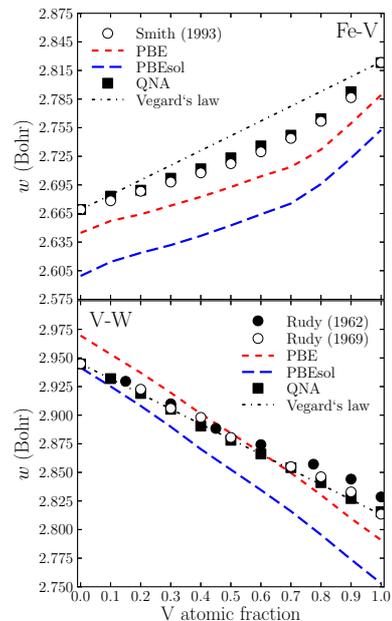}
 \caption{(Color online) Equilibrium Wigner-Seitz radii for Fe-V and V-W binary alloys. The present theoretical results (PBE/PBEsol: dashed lines, QNA: squares) are compare to Vegard's law (dashed-dotted line) and to experimental data (open and filled circles).\cite{62Rud1,69Rud2,93Smi1}}
 \label{figure1}
\end{figure}

According to Table \ref{table1}, neither PBE nor PBEsol yields systematic errors for the Wigner-Seitz radius of transition metals. Both of them underestimate the equilibrium volume of $3d$ metals (except for Cu) but overestimate those of the $5d$ metals. As a consequence, the PBE/PBEsol over-binding remains nearly the same when going from pure Fe to Fe-V solid solutions and then to pure V (Figure \ref{figure1}, upper panel). However, for the V-W binary alloy, the inconsistent errors for V and W result in a significantly larger theoretical (PBE/PBEsol) volume versus composition slope than the measured one (Figure \ref{figure1}, lower panel).

\begin{figure*}
 \includegraphics[width=0.65\textwidth]{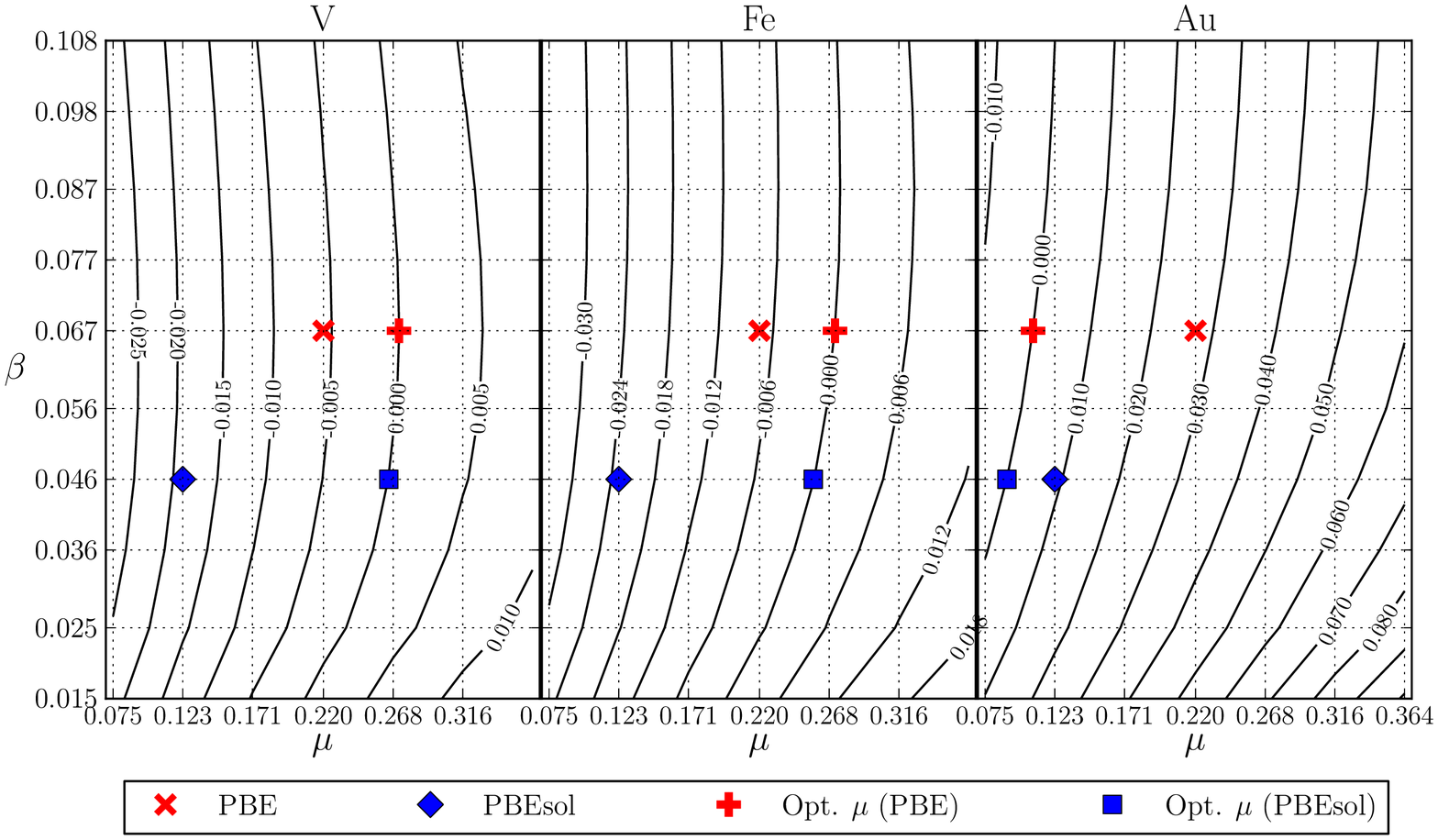}
 \caption{(Color online) Contours of relative errors in the Wigner-Seitz radius of V, Fe and Au. Specific sets of parameters corresponding to PBE, PBEsol, $\{\mu,\beta\}_{\rm opt}$ with $\beta=\beta_{\rm PBE}$ and $\beta=\beta_{\rm PBEsol}$ are shown by symbols. The dotted grid represents those $\{\mu,\beta\}$ pairs for which the calculations were performed. At every grid point, the equilibrium volume was derived from the total energy versus volume curves obtained using the particular $\{\mu,\beta\}$ parameters.}
 \label{figure2}
\end{figure*}

Many of the recent density functionals have been adjusted to model systems\cite{Kohn1998,Vitos2000,Armiento2005} or to metal surfaces.\cite{Perdew2008,Armiento2005} Following the same idea, we investigate whether it is possible to adjust a GDF so that it produces systematic errors for the bulk properties of metals. To this end, we consider the parameters $\mu$ and $\beta$ from the PBE/PBEsol scheme and compute the equilibrium volume of the selected mono-atomic metals as a function of these "variables" for $0.075\le \mu\le 0.364$ and $0.015\le\beta\le 0.108$. In figure \ref{figure2}, we show $w(\mu,\beta)$ for V, Fe and Au relative to the corresponding experimental values (Table \ref{table1}). We find that the errors in the theoretical $w$ change smoothly from negative to positive values as a function of $\mu$ but are non-monotonous in terms of $\beta$. As an consequence, changing $\beta_{\rm PBE}$ to $\beta_{\rm PBEsol}$, for instance, results in minor effect on the theoretical volume of V or Fe. Another interesting feature is the positive $\beta$ versus $\mu$ slope of the iso-error contour lines for $\beta\lesssim \beta_{\rm PBE/PBEsol}$. Hence, larger $\mu$ requires larger $\beta$ to preserve the error in the equilibrium volume. This is the often quoted "error cancelation" between exchange and correlation terms.

We find that for each element infinitely many pairs of $\{\mu,\beta\}$ yield vanishing error in the equilibrium volume. These combinations form a continuous line in the $\{\mu,\beta\}$-space (marked by 0.000 in Figure \ref{figure2}). The particular values $\mu_{\rm opt}$ corresponding to $\beta_{\rm opt}=\beta_{\rm PBE}$ and $\beta_{\rm opt}=\beta_{\rm PBEsol}$ are listed in Table \ref{table2}. We immediately observe that these "optimal" $\{\mu,\beta\}_{\rm opt}$ parameters are element specific. For instance, V and Fe require $\mu_{\rm opt} \sim 0.26-0.27$, whereas for the $5d$ elements $\mu_{\rm opt}$ drops below $\sim 0.13$. These demonstrate that it is not possible to define a unique pair of "optimal" $\{\mu,\beta\}$ parameters. In other words, at least within the PBE/PBEsol constraint, it is not possible to find a GDF that performs equally well for all metals.

\begin{table}[b]
\caption{\label{table2} Special values for $\mu_{\rm opt}$ corresponding to $\beta_{\rm opt}=\beta_{\rm PBE}$ and $\beta_{\rm opt}=\beta_{\rm PBEsol}$ for the seven selected metals.}
\begin{ruledtabular}
\begin{tabular}{llll}
Element & $\beta_{\rm PBE}$ & $\beta_{\rm PBEsol}$ \\ \hline
V & 0.2722 & 0.2646 \\
Fe & 0.2718 & 0.2570  \\
Cu & 0.1732 & 0.1587 \\
Nb & 0.1874 & 0.1750 \\
Pd & 0.1284 & 0.1124 \\
W & 0.1317 & 0.1185 \\
Au & 0.1080 & 0.0900
\end{tabular}
\end{ruledtabular}
\end{table}

In an attempt to find the "best" $\{\mu,\beta\}$ combination that leads to the smallest error in the equilibrium volume for the present set of mono-atomic solids, we minimized the cost function $C(\mu,\beta)\equiv \sum_{i} \left[w_i(\mu,\beta)/w_i(\rm expt)-1\right]^2$ ($i$ runs over the seven selected solids). Extending the search to $0\leq\mu\leq0.400$ and $0\leq\beta\leq0.260$ resulted in $\mu_{\rm g-opt} = 0.151990$ and $\beta_{\rm g-opt}=0.230019$ corresponding to $C_{\rm g-opt} = 0.0004$. In terms of the cost function, the above "globally" optimized pair of parameters represents marginal improvement over $C(\rm PBE) = 0.0012$ and $C(\rm PBEsol) = 0.0008$.

\begin{figure}
\includegraphics[width=0.35\textwidth]{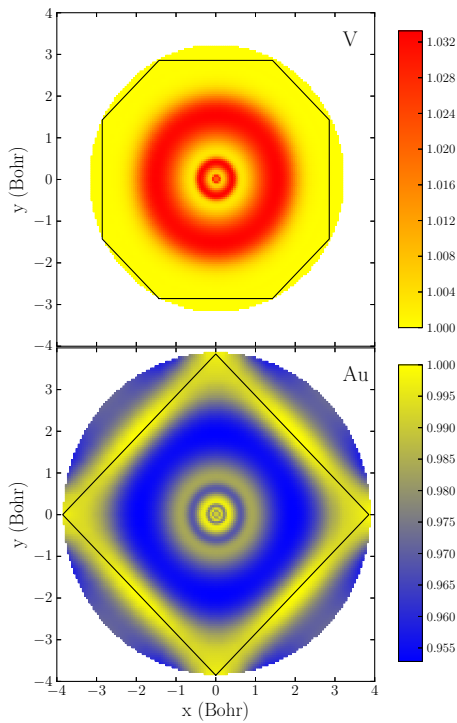}
\caption{(Color online) Contour plot of the $F_{xc}^{\rm opt}(r_s,s)/F_{xc}^{\rm PBE}(r_s,s)$ ratio within the (001) plane of the bcc V and fcc Au. The atoms are located at the origin ($x=y=0$), and the plots include points inside the sphere circumscribed to the Wigner-Seitz cells. The cross section between the cell boundary and the (001) plane is marked by thin solid lines.}
\label{figure3}
\end{figure}

In order to reveal the origin of the difference between the results obtained with PBE/PBEsol and those using $\{\mu,\beta\}_{\rm opt}$ (leading to vanishing errors in $w$), we monitor the ratio between the corresponding enhancement functions $F_{xc}^{\rm opt}(r_s,s)/F_{xc}^{\rm PBE}(r_s,s)$. Figure \ref{figure3} displays the above ratio within the (001) plane of bcc V and fcc Au. We notice that for both systems $F_{xc}^{\rm opt}(r_s,s)$ and $F_{xc}^{\rm PBE}(r_s,s)$ are close to each other around the cell boundary but show large (positive for V and negative for Au) deviations for points inside the atoms. That is, the interstitial region corresponds to nearly LDA regime, whereas the valence-core overlap region is where the details of the GDF become important. This observation is in line with those discussed by Fuchs \emph{et al.} \cite{Fuchs1998} and Csonka \emph{et al.} \cite{Csonka2009}

We conclude that for metals the primary error of a GDF is of local nature: the truly gradient-sensitive region is always localized around the atomic sites and the region in between the sites is less sensitive to the details of the density functional approximation. This observation opens the door to an alternative SFA. According to that, for a solid the optimal subsystems are the individual (element specific) valence-core overlap regions which are connected by the nearly homogeneous valence electron sea. For multi-component alloys, a possible realization of such \emph{quasi-non-local gradient-level approximation} is the superposition of the component-optimized gradient-level functionals. Mathematically a QNA functional may be expressed as
\begin{equation} \label{eq:Exc_qna}
E^{\rm QNA}_{\rm xc}[n] = \sum_q\int_{\Omega_q} {\rm d}^3r \epsilon^{\rm LDA}_x(n)F_{xc}^{{\rm opt}_q}(r_s,s),
\end{equation}
where $\Omega_q$ represents the Wigner-Seitz cell around atom $q$ and $F_{xc}^{{\rm opt}_q}$ is the PBE/PBEsol enhancement function based on $\{\mu,\beta\}_{{\rm opt}_q}$ optimized for the alloy component $q$. Since for any $\{\mu,\beta\}_{{\rm opt}_q}$, $F_{xc}^{{\rm opt}_q}(r_s,s)$ reduces to $\sim F_{xc}^{\rm LDA}(r_s)$ within the interstitial region (characterized by $s\rightarrow 0$), to a good approximation the kernel of the functional from (\ref{eq:Exc_qna}) is continuous at the cell boundaries.

We demonstrate the above QNA scheme in the case of bcc V-Fe and V-W solid solutions and CuAu and Cu$_3$Au intermetallic compounds adopting the L$1_0$ and L$1_2$ structures, respectively. Using $\{\mu,\beta\}_{\rm opt}$ obtained\cite{optmubeta} for bcc V, Fe and W, we computed the equilibrium Wigner-Seitz radii of V-Fe and V-W binary alloys as a function of chemical composition. The nearly perfect agreement between the present theoretical results (denoted by QNA in Figure \ref{figure1}) and the measured equilibrium radii shows that the proposed component-optimized approximation performs well for both dilute and concentrated alloys. For CuAu and Cu$_3$Au the PBE/PBEsol errors in $w$ relative to the experimental values\cite{Pearson1964} are $1.6/0.3\%$ and $1.1/0.7\%$, respectively. Using QNA with $\{\mu,\beta\}_{\rm opt}$ optimized for fcc Cu and Au, the above errors drop below $0.1\%$.

It has been found that although for each elemental metallic solid there are several optimal GDFs that give vanishing errors in the theoretical equilibrium volume, no globally accurate GDF exists. Starting from the concept of subsystem functional approach, we have introduced a quasi-non-local gradient-level approximation based on the component-optimized GDFs. The so constructed QNA has been shown to perform equally well for mono-atomic and multi-component metallic systems. The proposed scheme can easily be implemented in any density functional method, including pseudo-potential methods via properly designed pseudo-potentials. Finally we should mention that QNA can be turned fully \emph{ab initio} by using, instead the experimental equilibrium volumes, data provided by accurate many-body solvers.

\emph{Acknowledgements} The computer resources of the Finnish IT Center for Science (CSC) and Mgrid project and the financial support by the Swedish Research Council, the European Research Council (Grants No. 228074), the Swedish Foundation for International Cooperation in Research and Higher Education, the Hungarian Scientific Research Fund (research project OTKA 84078) and the G\"oran Gustafsson Foundation are acknowledged.


\begin{thebibliography}{}
\bibitem{Hohenberg1964} P. Hohenberg and W. Kohn, Phys. Rev. \textbf{136}, B864 (1964).
\bibitem{Kohn1965} W. Kohn and L. J. Sham, Phys. Rev. {\bf 140}, A1133 (1965).
\bibitem{Perdew1986} J. P. Perdew and Y. Wang, \prb {\bf 33}, 8800 (1986); J. P. Perdew, \prb {\bf 33}, 8822 (1986).
\bibitem{Perdew1996} J. P. Perdew, K. Burke, and M. Ernzerhof, \prl {\bf 77}, 3865 (1996).
\bibitem{Zhang1998} Y. Zhang and W. Yang, \prl {\bf 80}, 890 (1998).
\bibitem{Perdew2008}J. P. Perdew, A. Ruzsinszky, G. I. Csonka, O. A. Vydrov, G. E. Scuseria, L. A. Constantin, X. Zhou, and K. Burke, \prl {\bf 100}, 136406 (2008).
\bibitem{Armiento2002} R. Armiento and A.E. Mattsson, \prb \textbf{66}, 165117 (2002).
\bibitem{Kohn1996} W. Kohn, \prl \textbf{76}, 3168 (1996).
\bibitem{Kohn1998} W. Kohn and A. E. Mattsson, \prl {\bf 81}, 3487 (1998).
\bibitem{Vitos2000} L. Vitos, B. Johansson, J. Koll\'ar, and H. L. Skriver, \prb \textbf{62}, 10046 (2000); \emph{Ibid.} \pra \textbf{61}, 052511 (2000).
\bibitem{Armiento2005} R. Armiento and A.E. Mattsson, \prb \textbf{72}, 085108 (2005).
\bibitem{Constantin2009} L. A. Constantin, A. Ruzsinszky, and J. P. Perdew, \prb \textbf{80}, 035125 (2009).
\bibitem{Delczeg2011} L. Delczeg, E. K. Delczeg-Czirjak, B. Johansson, L. Vitos,
Journal of Physics: Condensed Matter \textbf{23}, 045006 (2011).
\bibitem{Tao2003} J. Tao, J.P. Perdew, V.N. Staroverov, and G.E.
Scuseria, \prl \textbf{91}, 146401 (2003).
\bibitem{Andersen1994} O. K. Andersen, O. Jepsen, and G. Krier,
in {\it Lectures on Methods of Electronic Structure Calculations},
ed. V. Kumar, O. K. Andersen, and A. Mookerjee, World Scientific
Publishing Co., Singapore, p. 63, (1994).
\bibitem{Vitos2000a} L. Vitos, H. L. Skriver, B. Johansson, and
J. Koll\'ar, Comp.\ Mat.\ Sci. {\bf 18}, 24 (2000).
\bibitem{Vitos2001} L. Vitos, \prb {\bf 64}, 014107, (2001); L. Vitos, I. A. Abrikosov, and B. Johansson, \prl {\bf 87}, 156401 (2001).
\bibitem{Kissavos2007} A. E. Kissavos, L. Vitos, and I. A. Abrikosov, \prb {\bf 75}, 115117 (2007).
\bibitem{Soven1967} P. Soven, Phys. Rev. {\bf 156}, 809 (1967); B. L. Gy\"orffy, \prb {\bf 5}, 2382 (1972).
\bibitem{MorseEOS} V. L. Moruzzi, J. F. Janak, and K. Schwarz, \prb {\bf 37}, 790 (1988).
\bibitem{Asato1999} M. Asato, A. Settels, T. Hoshino, T. Asada, S. Bl\"ugel, R. Zeller, and P. H. Dederichs, \prb {\bf 60}, 5202 (1999).
\bibitem{Ropo08} M. Ropo, K. Kokko, and L. Vitos, \prb {\bf 77}, 195445 (2008).
\bibitem{Staroverov2004} V. N. Staroverov, G. E. Scuseria, J. Tao, and
J. P. Perdew, \prb {\bf 69}, 075102 (2004).
\bibitem{Jiang2003} D. E. Jiang, and E. A. Carter, \prb {\bf 67}, 214103 (2003).
\bibitem{Mattsson2008} A. E. Mattsson, R. Armiento, J. Paier, G. Kresse,
J. M. Wills, and T. R. Mattsson, J. Chem. Phys. {\bf 128}, 084714 (2008).
\bibitem{Wyckoff1963} R. W. G. Wyckoff, \emph{Crystal Structures}, 2nd ed., (Interscience,New York, 1963).
\bibitem{Kaye1993} G. W. C. Kaye, and T. H. Laby, in \emph{Tables of physical and chemical constants}, 15th edition, (Longman, London, 1993).
\bibitem{Ho1974} C. Y. Ho, R. W. Powell, and P. E. Liley, in \emph{Thermal conductivity of the elements: A comprehensive review}, J. Phys. Chem. Ref. Data, vol. 3, pp. I-1 - I-796, (1974).
\bibitem{62Rud1} E. Rudy, F. Benesovsky, Monatsh. Chem. \textbf{93}, 693 (1962).
\bibitem{69Rud2} E. Rudy, \emph{Compendium of Phase Diagrams Data}, Air Force Materials Lab., Wright-Patterson Air Force Base, OH, Rep. No. AFML-TR-65-2, Part V (1969) 121
\bibitem{93Smi1} J. F. Smith, \emph{Phase Diagrams of Binary Iron Alloys}, (Materials Information Soc., Ohio, 1993).
\bibitem{Fuchs1998} M. Fuchs, M. Bockstedte, E. Pehlke, and M. Scheffler, \prb \textbf{57}, 2134 (1998).
\bibitem{Csonka2009} G. I. Csonka, J. P. Perdew, A. Ruzsinszky, P. H. T. Philipsen, S. Lebegue,
J. Paier, O. A. Vydrov, and J. G. \'Angy\'an, \prb \textbf{79}, 155107 (2009).
\bibitem{optmubeta} In these tests, for each element $\{\mu,\beta\}_{opt}$ was optimized using the room-temperature experimental volume since no 0 K data is available for alloys.
\bibitem{Pearson1964} W. B. Pearson, A Handbook of Lattice Spacings and Structures of Metals and Alloys, Pergamon, Oxford, vol. 4, (1964).
\end{thebibliography}
\end{document}